\documentclass[preprint2,numberedappendix]{emulateapj-rtx4}

\usepackage{amssymb,bm}
\usepackage{epsfig}
\usepackage{natbib}

\newcommand{\beq}{\begin{eqnarray}}
\newcommand{\eeq}{\end{eqnarray}}

\shorttitle{Constraining primordial magnetic fields}
\shortauthors{Kahniashvli et al.}

\begin{document}

\title{Constraining primordial magnetic fields through large scale structure}

\author{
Tina Kahniashvili$^{1,2,3}$,
Yurii Maravin$^{4}$,
Aravind Natarajan$^{1}$,
Nicholas Battaglia$^{1}$,
and \\
Alexander G. Tevzadze$^{5}$,
\email{tinatin@andrew.cmu.edu
($ $Revision: 1.55 $ $)}
}

\affil{
$^1$McWilliams Center for Cosmology and Department of Physics, Carnegie Mellon University,
    5000 Forbes Ave, Pittsburgh, PA 15213\\
$^2$Department of Physics, Laurentian, University, Ramsey Lake Road, Sudbury, ON P3E 2C, Canada\\
$^3$Abastumani Astrophysical Observatory, Ilia State University,
    3-5 Cholokashvili Ave., Tbilisi, 0194, Georgia\\
$^4$Department of Physics, Kansas State University, 116 Cardwell Hall, Manhattan, KS 66506\\
$^5$Faculty of Exact and Natural Sciences, Javakhishvili Tbilisi
State University, 3 Chavchavadze Ave., Tbilisi, 0128, Georgia }

\begin{abstract}
  We study primordial magnetic field effects on the matter
  perturbations in the Universe. We assume magnetic field generation prior
  to the big bang nucleosynthesis (BBN), i.e. during the radiation dominated
  epoch of the Universe expansion, but do not limit analysis by
  considering a particular magnetogenesis scenario.
  Contrary to previous studies, we limit the total magnetic field energy
  density and not the smoothed amplitude of the magnetic field at large
  (order of 1 Mpc) scales. We review several cosmological signatures, such as
  halos abundance, thermal Sunyaev Zel'dovich (tSZ) effect, and
  Lyman-$\alpha$ data. For a cross check we compare our limits with that
  obtained through the CMB faraday rotation effect and BBN.
  The limits are ranging between 1.5 nG and 4.5 nG
  for $n_B \in (-3; -1.5)$.
\end{abstract}


\maketitle

\section{Introduction}

Observations show that galaxies have magnetic fields with a
component that is coherent over a large fraction of the galaxy with
field strength of order $10^{-6}$ Gauss (G)
\citep{Beck,Widrow,Vallee}. These fields are supposed to be the
result of amplification of initial weak seed fields of unknown
nature. A recent study, based on the correlation of Faraday
rotation measures and MgII absorption lines (which trace halos of
galaxies), indicates that coherent $\mu$G-strength magnetic fields
were already in place in normal galaxies (like the Milky Way) when
the universe was less than half its present age~\citep{kronbergetal08}.
This places strong constraints both on
the strength of the initial magnetic seed field and the time-scale
required for amplification. Understanding the origin and evolution
of these fields is one of the challenging questions of modern
astrophysics. There are two generation scenarios under
discussion currently:
a bottom-up (astrophysical) one, where the needed seed
field is generated on smaller scales; and, a top-down (cosmological)
scenario, where the seed field is generated prior to galaxy formation
in the early universe on scales that are large now. More precisely,
astrophysical seed field sources include battery mechanisms, plasma
processes, or simple transport of magnetic flux from compact systems
(e.g.\ stars, AGNs), where magnetic field generation can be extremely
fast because of the rapid rotation \citep{Kulsrud:2007an}. Obviously,
the correlation length of such a seed field cannot be larger than a
characteristic galactic length scale, and is typically much smaller.
In the cosmological seed field scenario, \citep{Kandus:2010nw}, the
seed field correlation length could be significantly larger than the
current Hubble radius, if it was generated by quantum fluctuations
during inflation. There are different options for seed field
amplification, ranging from the MHD dynamo to the adiabatic
compression of the magnetic field lines during structure formation
\citep{Beck}. The presence of turbulence in cosmic plasma plays a
crucial role in both of these processes. The MHD turbulence was
investigated a long time ago when considering the processes in
astrophysical plasma, while there is a lack of studies when
addressing the turbulence effects in cosmological contexts
\citep{B03}. In the late stages of evolution the energy
density present in the form of turbulent motions in clusters can be
as large as 5-10\% of the thermal energy density \citep{kravtsov}.
This can influence the physics of clusters
\citep{Subramanian:2005hf}, and/or at least should be modeled
correctly when performing large scale simulations \citep{LSS1,LSS2}.
The proper accounting of the MHD turbulence effects is still under
discussion \citep{LSS3}. Both astrophysical and primordial
turbulence might have distinctive observational signatures. As we
already noted above, the most direct signature of MHD turbulence is
the observed magnetic fields in clusters and galaxies.

Galactic magnetic fields are usually measured through the induced
Faraday rotation effect (see \cite{Vallee}) and, as mentioned above, the
coherent field magnitude is of order a few $\mu$G with a typical
coherence scale of 10 kpc.\footnote{On the other hand, simulations
starting from constant comoving magnetic fields of $10^{-11}$G show
clusters generating fields sufficiently large to explain Faraday
rotation measurements \cite{Dolag,jedamzik1}.} On larger scales
there have been recent claims of an observed lower limit of order
$10^{-15}-10^{-16}$ G on the intergalactic magnetic field
\citep{neronov,limit2,dolag2}, assuming a correlation length of
$\lambda\ge 1$~Mpc, or possibly two orders of magnitude smaller
\citep{dermer1}. An alternative approach to explain the blazar
spectra anomalies has been discussed by \cite{Broderick:2011ab},
where two beam plasma instabilities were considered.{\footnote{The
recent study \cite{zero} claims that proper accounting for
uncertainties of the source modeling leads to consistence with a
zero magnetic field hypothesis.}}  Although these instabilities are
well tested through numerical experiments for laboratory plasma for
a given set of parameters such as a temperature and energy densities
of beams and background, its efficiency might be questioned for
cosmological plasma because of a significantly different (several
orders of magnitudes) beam, and background temperature and energy
densities. Prior to these observations, the intergalactic magnetic
field was limited only to be smaller than a few nG from cosmological
observations, such as the limits on the cosmic microwave background
(CMB) radiation polarization plane rotation \citep{yamazaki} and on
the Faraday rotation of polarized emission from distant blazars and
quasars \citep{a2}.

In the present paper we consider the presence of a primordial magnetic
field in the Universe and give a simplified description of its
effect on  large scale structure formation.
We assume that the magnetic field has been generated during the radiation
dominated epoch and prior to  big bang nucleosynthesis (BBN).
Since the magnetic energy density contributes to the relativistic component, the presence of such a magnetic field affects the moment of  matter-radiation
equality, shifting it to a later stages.
We focus
on the linear matter power spectrum in order to show that even if the
total energy density present in the magnetic field (and as a
consequence in magnetized turbulence) is small enough, its effects
might be substantial, and the effect becomes stronger due to
non-linearity of processes under consideration.

It has become conventional to derive the cosmological effects of a
seed magnetic field by using its spectral shape (parameterized by the
spectral index $n_B$) and the smoothed value of the magnetic field
($B_\lambda$) at a given scale $\lambda$ (which is usually taken to
be 1 Mpc). In \cite{ktr09} we developed a different and more
adequate formalism based on the effective magnetic field value that
is determined by the total energy density of the magnetic field.
Such an approach has been mostly motivated by the simplest energy
constraint on the magnetic field generated in the early universe.
In order to preserve BBN physics, only 10\% of the
relativistic energy density can be added to the radiation energy density,
leading to the limit on the total magnetic field energy density
corresponding to the effective magnetic field value order of $10^{-6}$ G.
More precise studies of the influence of the primordial magnetic field on
the expansion rate and the abundance of light elements
performed recently \citep{Yamazaki:2012jd,Kawasaki:2012va}, lead to
effective magnetic field amplitudes with order of $1.5 -1.9 \times 10^{-6}$ G.

The described formalism has been applied to describe two different effects of
the primordial magnetic field; the CMB Faraday rotation effect and
mass dispersion \citep{ktspr10}. As a striking consequence, we show
that even an extremely small smoothed magnetic field of $10^{-29}$ G
at 1 Mpc, with the Batchelor spectral shape ($n_B=2$) at large
scales, can leave detectable signatures in CMB or LSS statistics. In
the present investigation we focus on the thermal Sunyaev-Zel'dovich
effect, the cluster number density, and Lyman-$\alpha$ data. The large scale
based tests such as tSZ, Lyman-$\alpha$, cosmic shear (gravitational lensing),
X-rays cluster surveys, have been studied in \cite{shaw,tashiro,tashiro2,f12,pandey},
but again in the context of a smoothed magnetic field.
Another possible observational signature of  large-scale correlated
cosmological magnetic fields may be found in cosmic ray acceleration,
and corresponding gamma ray signals, (see Ref. \citep{kusenko} and references therein).
These observational signatures of the primordial magnetic field are beyond
the scope of the present paper.
We also do a more precise data analysis,
and we do not focus only on  inflation-generated magnetic fields.

The structure of the paper is as follows. In Sec. II we briefly
review the effective magnetic field formalism and discuss
the effect on the density perturbations. In Sec. III we review
observational consequences and derive the limits on primordial
magnetic fields. Conclusions are given in Sec. IV.

\section{Modeling the Magnetic Field Induced Matter Power Spectrum}
We assume that the primordial magnetic field has been generated
during or prior to BBN, i.e., well during the radiation dominated
epoch.\footnote{Note that some results of this paper can be applied
also to the case when magnetic fields are generated during the
matter dominated epoch, but with several "caveats": in this case the
BBN limits will not be valid, since the magnetic field will not be
present during matter-radiation equality and will not affect the
expansion rate of the early universe and light element abundances.
On the other hand, if the magnetic field has been generated prior to
recombination, the CMB limits must be used. For any other field
generated before  reionization and first structure formation only
the large-scale structure tests may apply. We thank the anonymous
referee for pointing out this issue. }
A stochastic Gaussian magnetic field is fully described by its
two-point correlation function. For simplicity, we consider  the
case of a non-helical magnetic field\footnote{We limit ourselves  to
 considering a non-helical magnetic field because the density
perturbations, and as a result the matter power spectrum, is not
affected by the presence of magnetic helicity.},  for which the
two-point correlation function in wavenumber space is \citep{ktspr10}
\begin{equation}
\langle B^\star_i({\mathbf k})B_j({\mathbf k'})\rangle =(2\pi)^3
\delta^{(3)} ({\mathbf k}-{\mathbf k'}) P_{ij}({\mathbf{\hat k}})
P_B(k). \label{spectrum}
\end{equation}
Here, $i$ and $j$ are spatial indices; $i,j \in (1,2,3)$,
$\hat{k}_i=k_i/k$  is a unit wavevector; $P_{ij}({\mathbf{\hat
k}})=\delta_{ij}-\hat{k}_i\hat{k}_j$ is the transverse plane projector;
$\delta^{(3)}({\mathbf k}-{\mathbf k'})$ is the Dirac delta function,
and $P_B(k)$ is the power spectrum of the magnetic field.

The smoothed magnetic field $B_\lambda$ is defined through the mean-square
magnetic field, ${B_\lambda}^2 = \langle {\mathbf B}({\mathbf x})
\cdot {\mathbf B}({\mathbf x})\rangle |_\lambda $, where the
smoothing is done on a comoving length $\lambda$ with a Gaussian
smoothing kernel function $\propto \mbox{exp}[-x^2/\lambda^2]$.
Corresponding to the smoothing length $\lambda$ is the smoothing
wavenumber $k_\lambda=2\pi/\lambda$. The power spectrum $P_B(k)$ is
assumed to depend on $k$ as a simple power law function on large scales,
$k<k_D$ (where $k_D$ is the cutoff wavenumber),
\begin{equation}
P_B(k) = P_{B0}k^{n_B}= \frac{2\pi^2 \lambda^3
B^2_\lambda}{\Gamma(n_B/2+3/2)} (\lambda k)^{n_B},
\label{energy-spectrum-H}
\end{equation}
and assumed to vanish on small scales where $k>k_D$.

We define the effective magnetic field $B_{\rm eff}$ through the
magnetic energy density $\rho_B = {B_{\rm eff}^2}/({8\pi})$. In
terms of the smoothed field, the magnetic energy density is given by
\begin{equation}
{\rho}_B(\eta_0) = \frac{B_\lambda^2 (k_D \lambda)^{n_B+3}}{8\pi
\Gamma(n_B/2+5/2)} ~,
\end{equation}
and thus $B_{\rm eff} = B_\lambda (k_D \lambda)^{(n_B
+3)/2}/\sqrt{\Gamma(n_B/2+5/2)}$. For the scale-invariant spectrum
$n_B = -3$ and $B_{\rm eff} = B_\lambda$ for
all values of $\lambda$. The scale-invariant spectrum is the only case
where the values of the effective and smoothed fields coincide. For
causal magnetic fields with $n_B =2$ the smoothed magnetic field
value is extremely small for moderate values of the magnetic
field.

We also need to determine the cut-off scale $k_D$. We assume that
the cut-off scale is determined by the Alfv\'en wave damping scale
$k_D \sim v_A L_S$, where $v_A$ is the Alfv\'en velocity and $L_S$ is
the Silk damping scale \citep{jedamzik98,sb98}. Such a description
is more appropriate when  dealing with a homogeneous magnetic
field, and the Alfv\'en waves are the fluctuations of ${\bf B}_1({\bf
x})$ with respect to a background homogeneous magnetic field ${\bf
B}_0$ ($|{\bf B}_1 |\ll |{\bf B_0}|$). In the case of a stochastic
magnetic field we generalize the Alfv\'en velocity definition from
\cite{mkk02}, by referring to the analogy between the effective
magnetic field and the homogeneous magnetic field. Assuming that the
Alfv\'en velocity is determined by $B_{\rm eff}$, a simple
computation gives the expression of $k_D$ in terms of $B_{\rm eff}$:
\begin{equation}
\frac{k_D}{1{\rm Mpc}^{-1}} = 1.4 \sqrt{\frac{(2\pi)^{n_B+3}
h}{\Gamma(n_B/2+5/2)}} \left(\frac{10^{-7}{\rm G}}{B_{\rm
eff}}\right). \label{rho1}
\end{equation}
Here $h$ is the Hubble constant in units of 100 km s$^{-1}$
Mpc$^{-1}$.

Note that any primordial magnetic field
generated prior or during BBN should satisfy the BBN bound
(for a recent studies of  primordial magnetic fields effects on
 BBN processes and corresponding limits see
\citep{Yamazaki:2012jd,Kawasaki:2012va}).
Assuming that
the magnetic field energy density is not damped away by MHD
processes, the BBN limit on the effective magnetic field strength,
$B_{\rm eff} \leq 1.5 - 1.9  \times 10^{-6}$ G, while transferred in terms
of $B_\lambda$ the BBN bounds results in extremely small values for
causal fields, see \citep{cd01,ktr09}.

The primordial magnetic field affects all three kinds of metric
perturbations, scalar (density), vector (vorticity), and tensor
(gravitational waves) modes through the Einstein equations. The
primordial magnetic field generates a matter perturbation power
spectrum with a different shape compared to the standard
$\Lambda$CDM model. As we noted above in this paper we focus on
matter perturbations. As it has been shown by \citep{kim,gs03}, the
magnetic-field-induced matter power spectrum $P(k)
\propto k^4$ for $n_B > -1.5$ and $\propto k^{2n_B+7}$ for $n_B \le
-1.5$.
\begin{figure}[t]
\begin{center}
\includegraphics[width=0.99\columnwidth]{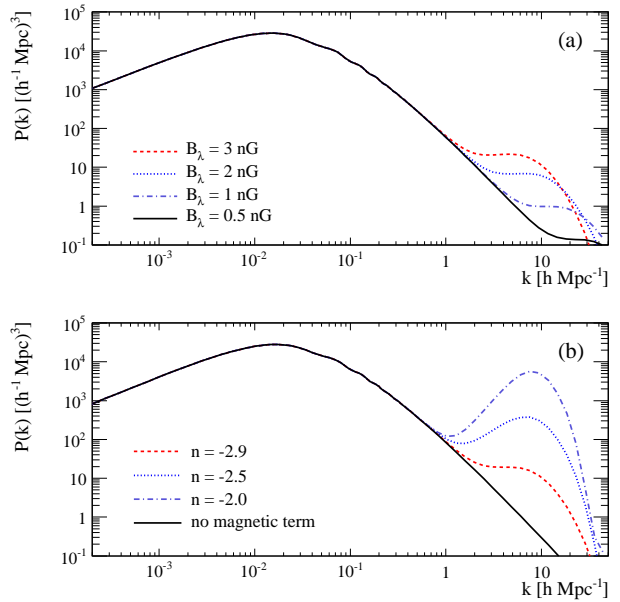}
\end{center}\caption[]{
The magnetic field matter power spectra for $n_B=-2.9$ and for different values of $B_\lambda$ (a) and for $B_\lambda=3$ nG and for different
values of $n_B$ (b).}
\label{matter1}
\end{figure}
This in turn affects the formation of rare objects like galaxy
clusters which sample the exponential tail of the mass function.
\cite{shaw} study in great detail the formation of the
magnetic field matter power spectrum through analytical description,
and provide a modified version of CAMB that includes the
possibility of a non-zero magnetic field. We have used the CAMB code to
determine the matter power spectra for a wide range of the magnetic
field amplitudes and spectral indices. These spectra are shown in
Fig.~(\ref{matter1}). It is obvious that the
matter power spectrum is sensitive to the values of the cosmological
parameters: the Hubble constant in  units of 100 km$/{\rm sec}/$Mpc,
$h$, $\Omega_M$, and $\Omega_b$, as well as  the density parameter
of each  dark matter component, i.e., $\Omega_{\rm {cdm}}$ and
$\Omega_\nu$ (here, $M$, $b$, ${\rm cdm}$, and $\nu$ indices refer to
matter, baryons, cold dark matter, and neutrinos respectively, and
$\Omega$ is the density parameter. To generate the matter plot we
assume the standard flat $\Lambda$CDM model with zero curvature, and
we use the following cosmological parameters: $\Omega_b h^2 =0.022$,
$\Omega_C h^2 = 0.1125$, and $h=0.71$. For simplicity, we assume
massless neutrinos with three generations.\footnote{The standard $\Lambda$CDM
model matter power spectrum $P_{\Lambda CDM} (k)$ assumes a close to
scale-invariant (Harrison-Peebles-Yu-Zel'dovich) post-inflation
energy density perturbation power spectrum $P_0(k) \propto k^n$,
with $n \sim 1$. } As we can see the increase of the smoothed field
amplitude results in the additional power spectrum shift to the
left, while increasing the value of $n_B$ makes the vertical
shift.
As we can see the large-scale tail (small wavenumbers) of the matter
power spectrum is unaffected by the presence of the magnetic field. Below
we address some of effects induced by the presence of the magnetic field,
especially on large scales.

\section{Observational Signatures}

Primordial magnetic fields can play a potentially important role in
the formation of the first large-scale structures.

\subsection{The Thermal Sunyaev-Zel'dovich effect}

As demonstrated in \cite{shaw,tashiro,Paoletti:2012bb} the strength of the primordial
magnetic field affects the growth of structure. The power spectrum
of secondary anisotropies in the CMB caused by the thermal
Sunyaev-Zel'dovich effect (tSZ) is a highly sensitive probe of the
growth of structure \citep[e.g.][]{Komatsu:2002wc}. The tSZ angular power
spectrum probes the distribution of galaxy clusters on the sky
essentially out to any redshift. At $l \simeq 3000$, half of the
contribution to the SZ power spectrum comes from matter halos with
masses greater than $\sim 2\times10^{14} M_{\sun}$ at redshifts
less than $z \simeq 0.5$, see \cite{battaglia:2011,Trac:2010sp}.
\begin{figure}[t]
\begin{center}
\includegraphics[width=\columnwidth]{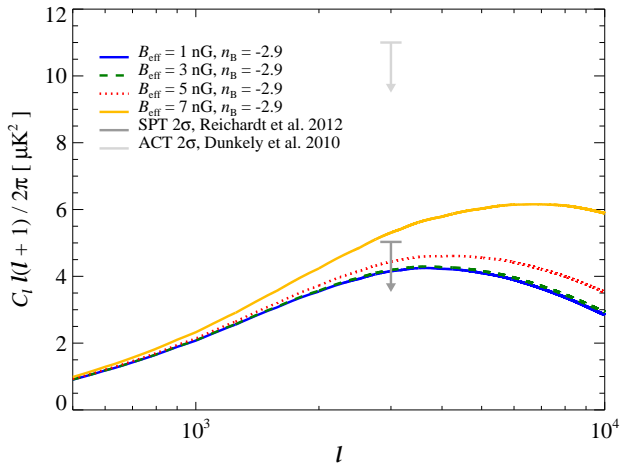}
\end{center}\caption[]{
The tSZ power spectrum predictions at 150 GHz varying the
primordial magnetic field model at fixed cosmological parameters,
most importantly  $\sigma_8 = 0.8$. These predictions are compared
against the recent upper limits from ACT \citep{Dunkley:2010ge} and
SPT \citep{Reichardt:2011yv} at $\ell = 3000$. The current upper limits on the tSZ
amplitude at $\ell = 3000$ do not constrain the primordial magnetic
field parameters $B_{\mathrm{eff}}$ and $n_{\mathrm{B}}$ as well as
other observations.}\label{tsz}
\end{figure}

All the previous work on how primordial magnetic fields affect
the tSZ power spectrum have used the model from \cite{Komatsu:2002wc},
here referred to as KS model, which has been shown to be incompatible with
recent observations of clusters \citep{Arnaud:2009tt}
and tSZ power spectrum
measurements by \cite{Lueker:2009rx}. Using the KS model for primordial
magnetic field studies also ignores all the recent advancements in
tSZ power spectrum theory and predictions that illustrate the
importance of properly modeling the detailed astrophysics of the
intracluster medium
\citep[e.g.][]{Battaglia:2010tm,battaglia:2011,Shaw:2010mn,Trac:2010sp}. We
modify the code described in \cite{shaw} to include these improvements
by changing the pressure profile used in their model from KS to the
profile given in \cite{Battaglia:2010tm,battaglia:2011}. The results from
the new pressure profile are shown in Fig.~(\ref{tsz}) with the greatest difference
being the amplitude of the new tSZ power spectrum
is approximately two times lower than previous predictions and below
the current observational constraint from ACT \citep{Dunkley:2010ge}
and SPT \citep{Reichardt:2011yv} at $\ell = 3000$. Updating
the theory predictions for the tSZ power significantly reduces the
constraints put on primordial magnetic field parameters using these
observations. In Fig.~(\ref{tsz}) we illustrate that  magnetic fields with an effective
amplitude of order of 5 nG are almost excluded.
Given that there is additional uncertainty in the
theoretical modeling of the tSZ
\citep[e.g.][]{Battaglia:2010tm,battaglia:2011,Shaw:2010mn,Trac:2010sp},
combined with significant contributions from other secondary
sources \citep{Reichardt:2011yv,Dunkley:2010ge} around $\ell \sim
3000$, for example from dusty star forming galaxies,
future tSZ power spectrum measurements are not going to be
competitive in constraining primordial magnetic fields parameters.

\subsection{Halo Number Density}

The predicted halo number density $N_{\rm{pred}}(M>M_0, z)$
depends on the considered cosmological model. One of important
characteristics of a cosmological model is the  linear matter power
spectrum that we reviewed in Sec. II above. Below we discuss the halo
number count dependence on the presence of the magnetic field.

The halo mass function at a redshift $z$ is $N(M>M_0,
z)=\int_{M_0}^\infty dM ~n(M, z)$, where $n(M, z)dM$ is the comoving
number density of collapsed objects with mass lying in the interval
$(M, M+dM)$, and it can be expressed as
\begin{equation}
n(M, z) = \frac{2\rho_M}{M} \nu f(\nu) \frac{d\nu}{dM}.
\label{n1}
\end{equation}
The multiplicity function $\nu f(\nu)$  is a universal function of the peak height \citep{ps74}
$\nu = \delta_C/\sigma(R)$, where $\sigma(R,z)$ is the r.m.s. amplitude of density
fluctuations smoothed over a sphere of  radius
$R=(3M/4\pi\rho_M)^{1/3}$, and the critical density contrast $\delta_C \simeq 1.686$ is the density contrast for a linear
overdensity able to collapse at the redshift $z$. Here, $\rho_M$ is the mean matter density at the redshift $z$.
For gaussian  fluctuations
$
\nu f(\nu) \propto {\rm exp}[-\nu^{2}/2]$ \citep{ps74}, where the normalization constant is fixed by the requirement that
 all of the mass lie in a given halo $\int \nu f(\nu) d\nu = 1/2$ \citep{white2}.
The evolution of the halo mass function $n(M, z) $ is mostly determined by the $z$
dependence of $\sigma(R,z)$.

\begin{figure}[t]
\begin{center}
\includegraphics[width=\columnwidth]{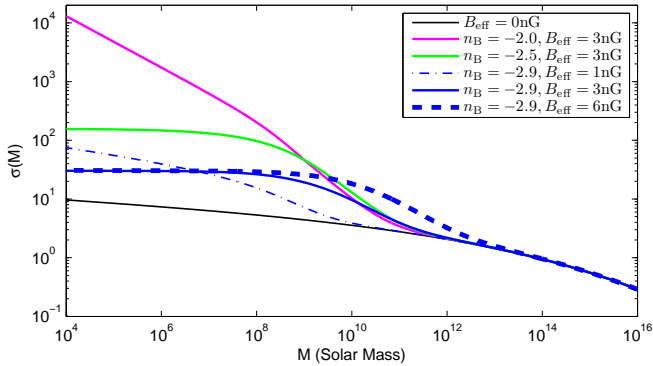}
\end{center}\caption[]{
$\sigma(M, z=0)$ for different effective magnetic field values $B_{\rm
eff}$ and spectral index $n_{\rm B}$.}\label{sigma0}
\end{figure}

The r.m.s amplitude of density fluctuations $\sigma^{2}(R,z)$ is related to
the linear matter power spectrum $P(k,z)$ through \citep{jenkins}
\begin{equation}
\sigma^{2}(R,z)= \frac{D(z)^2}{2 \pi^{2}}\int\limits_{0}^{\infty}
P(k,z)|W(kR)|^{2}k^{2}dk, \label{sigma}
\end{equation}
where $D(z)$  is the growth factor of linear perturbations normalized as  $D (z=0) = 1$ today,
$W(kR)$ is the Fourier transform of the top-hat window
function, $ W(x)={3}(\sin{x}- x\:\cos{x})/x^3 $.
In Fig.~\ref{sigma0} we illustrate the $\sigma(M, z=0)$ function for the different
values of the effective magnetic field, $B_{\rm eff}$,  and
the spectral index $n_{\rm B}$. The smaller amplitude of
the magnetic field results in modifications at smaller mass scales.
The $\sigma (M)$ dependence on the magnetic field characteristics
is also derived in \citep{ktspr10}, but contrary to the case
presented here, reflects {\it only} the $\sigma(M)$
induced by the pure magnetic field. In the present work we
derive the effect from the magnetic field on the overall matter
dispersion, including the standard density perturbations.
The value of $\sigma(M)$ at $M=2 \times 10^{14}M_{\rm Sun}$
is around 0.8 agreeing well with observational data, see \citep{vikhlinin}.

Numerical computation results for $n(M, z)$ are not accurately fit
by the PS expression $\nu f(\nu) \propto {\rm exp}[-\nu^2/2]$, see Refs. \citep{sheth,jenkins,hu-k}. Several
more accurate modifications of $n (M, z)$ have been proposed. Here,
we use the ST modification \cite{sheth}, as defined, (see Eq. 5 of Ref. \citep{white2})
\begin{eqnarray}
f(\nu)  \propto
[ 1+ (a\nu^2)^{-p}] (a\nu^2)^{-1/2}
\mbox{exp}[-a\nu^{2}/2] ~,
\label{f}
\end{eqnarray}
where the parameters $a=0.707$, and $p=0.303$  are fixed by fitting to
the numerical results \citep{white} (for the PS case: $a=1$ and $p=0$)
\cite{sheth}. With this choice of parameter values the mass of
collapsed objects in Eq.~(\ref{f})  must be defined using a fixed
over-density contrast with respect to the background density
$\rho_M$, and this requires accounting for the mass conversion
between $M_{180b}$ and $M_{200c}$. Such a conversion depends on
cosmological parameters, (see Fig.~1 of \cite{white}). Here, we use an
analytical extrapolation  of this figure to do the conversion for
$\Omega_M \in (0.2, 0.35)$.

The difference induced by the magnetic field in the matter power
spectrum $P(k)$ can potentially modify the $\delta_C$
parameter entering in Eq.~(\ref{f}), that will result in  different
halo number counts. On the other hand, here we focus on
the first order effects, so we neglect all changes induced by the
magnetic field in the Sheth-Tolmen model parameter fitting (see \cite{sheth}).
We also use the halo number count function at $z=0$ because
we are focusing only on the linear power spectrum, and all effects
related to the magnetic field non-linear evolution (see \cite{min})
during the structure formation are neglected. We will present a more
realistic scenario of the first object formation in future works.

\begin{figure}[t]
\begin{center}
\includegraphics[width=\columnwidth]{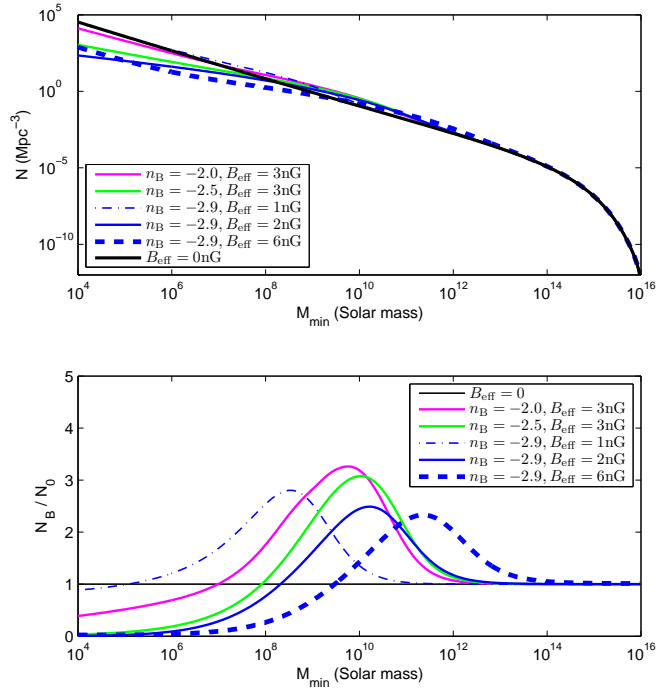}
\end{center}\caption[]{
Halo number density $N(M>M_0)$ (top panel) and
ratio of number density for magnetic and
non-magnetic simulations $N_{\rm B})/N_0$ (bottom panel)
for different effective magnetic field values $B_{\rm eff}$
and spectral index $n_{\rm B}$, and $z=0$. Number of small mass objects
($M\sim 10^4 M_\Sun$) in magnetized case can be reduced down by
factor of 100 compared to the non-magnetic number,
 object
number count excess occurs for objects with mass
around ($M\sim 10^{10} M_\Sun$).
}\label{Delta}
\end{figure}

In Fig.~(\ref{Delta}) (top panel) we illustrate the halo mass function
today ($z=0$) for different values of $B_{\rm eff}$ and $n_{\rm B}$.
As we can see, the magnetic field presence affects the small mass
ranges, reducing the abundance of low mass objects.
We do not present here any statistics using halo data accounting
for several uncertainties involving clusters physics \citep{battaglia:2011}.
On the other hand, we would like to underline that the presence of a high
enough magnetic field might be a possible explanation of the
low mass objects abundance, which is one of the unsolved puzzles in
$\Lambda$CDM cosmologies.

To get a better understanding of the magnetic field influence
on the halo abundance, we plot the ratio of halo number
density of $\Lambda$CDM models with and without magnetic fields
(see Fig. \ref{Delta}, bottom panel). In the high mass limit all
magnetized $\Lambda$CDM models compared to the $\Lambda$CDM model predict slightly (a relative
difference of the order of $10^{-5}$) higher halo number density.
Number density excess peaks around halos with mass
($M \sim 10^{10} M_\Sun$) and is strongly affected by the
effective magnetic field value, as well as on the spectral shape.
In contrast, at low mass limit $M < 10^7 M_\Sun$,  number of
objects can be significantly lower as then its non-magnetic value.

\subsection{Lyman-$\alpha$ data}

The small scale modifications induced by the primordial magnetic
field must be reflected in  first object formation in the Universe,
i.e., the objects at high redshifts. The
most important class of such objects are damped Lyman-$\alpha$
absorption systems.\footnote{These objects have a high column density
of neutral hydrogen ($N_{\rm HI}> 10^{20}$cm$^{-2}$) and are detected
by means of absorption lines in quasar spectra \citep{wolfe93}.
Observations at high redshift have lead to estimates of the abundance
of neutral hydrogen in damped Lyman-$\alpha$ systems \citep{wt}. The
standard view is that damped Lyman-$\alpha$ systems are a population of
protogalactic disks \citep{wolfe93}, with a minimum mass of
$M = 10^{10}h^{-1}M_{\rm Sun}$ \citep{H1995}.}
To describe these systems it is possible to use semi-analytical modeling.
Lyman-$\alpha$ systems has been used to constrain different cosmological
scenarios, see Ref. \citep{uros}, and references therein.
Lyman-$\alpha$ data is very sensitive to the matter power spectrum
around $k \simeq 10^{-1}-10^2$ Mpc$^{-1}$, wavenumbers that are affected
by the primordial magnetic field \citep{shaw}. As we will see below these
systems can be used to place stringent constraints on magnetic field
properties.

We do not go through the detailed modeling of Lyman-$\alpha$ systems,
leaving this for more precise computations, but we use the direct
comparison of the reconstructed matter power spectrum and the
theoretical matter power spectra affected by the primordial magnetic
field.

\begin{figure}[t]
\begin{center}
\includegraphics[width=0.99\columnwidth]{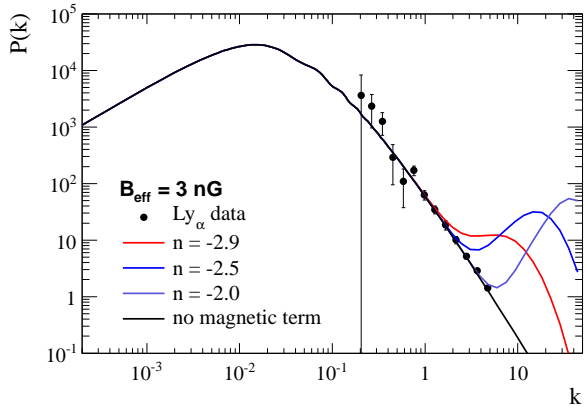}
\end{center}\caption[]{
The magnetic field matter power spectra for different values of
$n_B$ and data points from \cite{Croft} }\label{beff}
\end{figure}

For this study we use Lyman-$\alpha$ data obtained by the Keck
telescope~\citep{Croft}.
To get a conversion of data  points
(accounting that we use the wavevector $k$ units $h/$Mpc, we
multiply data by the conversion factor
$$ \frac{ 100 \; \sqrt{ \Omega_{\rm m} \left( 1 + z \right )^3
+ \Omega_\Lambda }} { 1 + z } $$ given in Ref.~\citep{kim}.
As the data is given at redshift 2.72, we translate the data to redshift zero
by multiplying it by the square of the ratio of the growth factor at redshift
zero to that at redshift 2.72. We compute the growth factors using the
{\sc ICOSMOS} calculator.\footnote{{\sc
ICOSMOS} Calculator is available at
http://www.icosmos.co.uk/index.html.} Thus, we multiply the data by 8.145 to estimate the
Lyman-α data at redshift z = 0. The comparison of the theoretical predicted
matter power spectrum and Lyman-$\alpha$ data is given in Fig. (\ref{beff}).

We use $\chi^2$ statistics to compare the predicted model with
Lyman-$\alpha$ data. We assume no correlation between the uncertainties
in the $P({\mathbf k})$ measurements for different ${\mathbf k}$ values
and find no evidence for primordial magnetic fields.

\begin{figure}[t]
\begin{center}
\includegraphics[width=0.99\columnwidth]{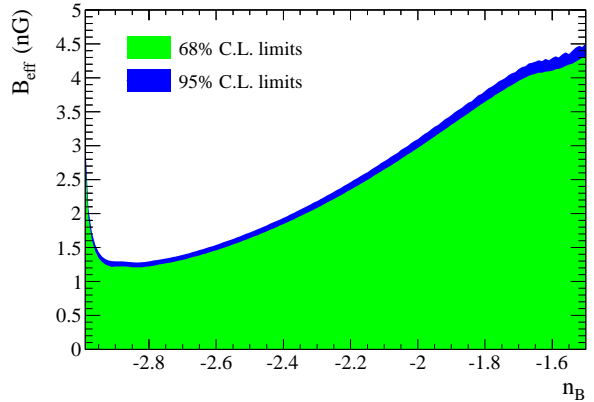}
\end{center}\caption[]{
The effective magnetic field limits from  Lyaman-$\alpha$ data  for different values of
$n_B$ }\label{limits}
\end{figure}

\begin{figure}[t]
\begin{center}
\includegraphics[width=0.99\columnwidth]{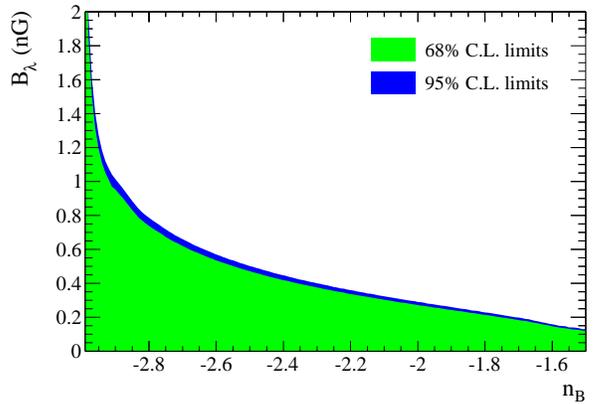}
\end{center}\caption[]{
The smoothed magnetic field limits from  Lyman-$\alpha$ for different values of
$n_B$ }\label{Blambdalimits}
\end{figure}

The 95\% and 68\% confidence level limits are given in Fig.
\ref{limits}. The limits on $B_\lambda$ are given Fig. \ref{Blambdalimits}.
We explicitly present the limits for $B_{\rm eff} $ and
$B_\lambda$ just to show that they have  different behaviors
when the spectral index is increasing. In terms of the total
energy density of the magnetic field the limits are weaker if
we are considering the redder spectra.
At this point the total energy density of the phase transition
generated magnetic field is almost unconstrained.

\subsection{The CMB Faraday Rotation effect}

As we have already noted above the primordial magnetic field induces CMB polarization Faraday rotation, and
for a homogeneous magnetic field the rotation angle is given by, \citep{kl96}
\begin{equation}
\alpha \simeq 1.6^\circ \left(
\frac{B_0}{1\, {\rm nG}}\right) \left( \frac{30\, {\rm GHz}}{\nu_0}
\right)^2, \label{a2}
\end{equation}
where $B_0$ is the amplitude of the magnetic field, and $\nu_0$ is the frequency of the CMB photons.
In the case of a stochastic magnetic field we have to determine the r.m.s. value of the rotation angle, $\alpha_{\rm rms}$,
and the corresponding expression in terms of the effective magnetic field is given in \citep{ktspr10},
being
\begin{eqnarray}
\alpha_{\rm rms} & \simeq & 0.14^\circ
\left(\frac{B_{\rm eff}}{1\, {\rm nG}}\right)   \left(
\frac{100 \, {\rm GHz}}{\nu_0} \right)^2 \frac{\sqrt{n_B +3}}{ (k_D
\eta_0)^{(n_B +3)/2}} \nonumber \\
& \times & \left[ \sum_{l=0}^{\infty} (2l+1)
  l(l+1) \int^{x_S}_0 dx \, x^{n_B}j^2_l(x) \right]^{1/2}\!\!\!.
\label{a11}
\end{eqnarray}
Here, $\eta_0$ is the present  value of conformal time, $j_l(x)$ is a Bessel function with argument $x=k\eta_0$,
and $x_S =k_S \eta_0$ where $k_S = 2~{\rm Mpc}^{-1}$ is the Silk
damping scale. In the case of an extreme magnetic field which just
satisfies the BBN bound, $k_D $ might become less than the Silk
damping scale. In this case the upper limit in the integral above
must be replaced by $x_D = k_D \eta_0$.
Note, that for $n_B \rightarrow -3$, Eq. (\ref{a11}) is reduced to Eq.
(\ref{a2}) (see for details Ref. \citep{ktspr10}.

\begin{figure}[t]
\begin{center}
\includegraphics[width=0.99\columnwidth]{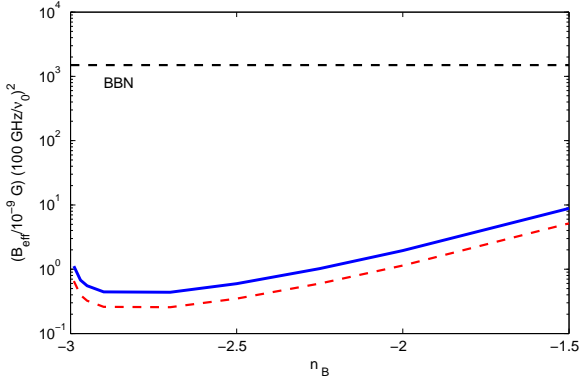}
\end{center}\caption[]{
The effective magnetic field values for different spectral index
$n_\mathrm{B}$. Solid and dashed lines correspond to the 95\% and 68\%
confidence levels, respectively. Upper limit set by BBN $B_\mathrm{eff} \sim {\mu {\rm G}}$, \citep{Yamazaki:2012jd,Kawasaki:2012va}, d
is shown by horizontal black line (BBN).}\label{Beff_nB}
\end{figure}

Here, we quote Ref. \citep{wmap} in order to determine the upper limits for the r.m.s. rotation angle.
Adding the statistical and systematic errors in quadrature
and averaging over WMAP \citep{wmap}, QUaD \citep{quad} and BICEP \citep{bi} (see for more details Ref. \cite{wmap}) with
 inverse variance weighting, the limits obtained were
$\alpha  = -0.25^0 \pm  0.58^0$ at (68\% CL), or
$-1.41^0 < \alpha < 0.91^0$ (95\% CL).
We obtain for the r.m.s. value (absolute) of the rotation angle
$|\alpha_{\rm rms.}| < 0.477^0$ and $|\alpha_{\rm rms} | < 0.997^0 $ (68\% C.L. and 95\% C.L., respectively)
assuming gaussian statistics.
In Fig.~(\ref{Beff_nB}) we display the upper limits of the effective magnetic field using the rotation angle constraints quoted above.
Note that these limits are an order of magnitude better than obtained previously in Ref. \citep{ktspr10} where we used the WMAP 7 year data alone.
For almost scale-invariant magnetic field the limits are around 0.5 nG. As we can see for $n_B > -0.5$ the BBN limits on the effective magnetic field strength are stronger than those coming from the CMB faraday rotation effect.
The situation is completely different when determining the limits for the smoothed magnetic field $B_{\lambda=1{\rm Mpc}}$ with $n_B >-2$ , which are extremely strong from BBN \citep{cd01,ktr09}, and moderate in the case of the large scale structures or the CMB birefrigence, see above.

\section{Conclusion}

In this paper we studied the large-scale signatures of cosmological
magnetic fields
generated during the radiation dominated epoch prior to the BBN.
We address such effects
as the thermal Sunyaev-Zel'dovich effect, halo number density, and
Lyman-$\alpha$ data. Due to several uncertainties present in tSZ and
halo abundance tests we find that Lyman-$\alpha$ measurements provide
the tightest constraints on the  primordial magnetic field energy density.
We express these limits in terms of the
effective value of the magnetic field, $B_{\rm eff}$. In the case of
the scale invariant spectrum $n_B=-3$ these limits are identical to
limits on the smoothed magnetic field $B_\lambda$, (smoothed over a
length scale $\lambda$ that is conventionally taken to be 1 Mpc).
For a steep magnetic field with spectral index $n_B = 2$ the difference
between the limits derived in terms of the effective and smoothed field
is several orders of magnitude. Also limits have different behavior
with increasing  $n_B$. At this point, as we underlined previously
\citep{ktspr10}
using the smoothed magnetic field can result in
some confusion:
the smoothed magnetic field at 1 Mpc scales is extremely small, while the total energy density of the magnetic field
is maximal allowed by BBN bounds
(see Refs.~\cite{Yamazaki:2012jd,Kawasaki:2012va} for more details on BBN bounds).
The small values of the magnetic fields for $n_B=2$ (that corresponds to the phase transition generated magnetic fields) might be treated as non-relevance on these fields. For example, in  Ref. \citep{shaw} it is claimed that  the magnetic field with
the spectral index greater than -2.5 is excluded \citep{shaw}, while as it is shown in Ref. \citep{ktr09}  the magnetic field with extremely small
smoothed field value $B_\lambda$ at $\lambda=1$ Mpc order of $10^{-29}$ Gauss with the spectral index $n_B=2$ can leave observable traces on the CMB and large scale structure formation.
The limits range between 1.5 nG and 4.5 nG
for $n_B \in (-3; -1.5)$.
These limits are comparable for those from the CMB polarization plane rotation.
Our results can be applied with some precautions to the primordial magnetic fields generated in the matter dominated epoch too, see sec. 2.

{\it Note} when this paper was in final stage of preparation
Ref. \citep{pandey} appeared showing that  magnetic fields
can be strongly constrained by  first object formation, in particular
through Lyman-$\alpha$ data.

We acknowledge useful comments from the anonymous referee.
We are greatly thankful to R. Shaw for useful comments and
discussion. The computation of the magnetic field power spectrum has
been performed using the modified version of CAMB, for details see
\citep{shaw}. We appreciate useful discussions with A.\ Brandenburg,
L. Campanelli, R. Croft, R. Durrer, A.\ Kosowsky, A.\ Kravtsov, F.
Miniati, K. Pandey, B. Ratra, U. Seljak, S. Sethi, and R. Sheth. We acknowledge partial
support from Swiss National Science Foundation SCOPES grant 128040,
NSF grants AST-1109180, NASA Astrophysics Theory Program grant
NNXlOAC85G. T.K.\ acknowledges the ICTP associate membership
program. A.N and N.B. are supported by a McWilliams Center for Cosmology
Postdoctoral Fellowship made possible by Bruce and Astrid McWilliams
Center for Cosmology. A.T.\ acknowledges the hospitality of the McWilliams Center
for Cosmology.

\newcommand{\yaraa}[3]{ #1, {ARA\&A,} {#2}, #3}
\newcommand{\yjour}[4]{ #1, {#2}, {#3}, #4}

\end{document}